# COMPACT RADIO CORES IN SEYFERT GALAXIES

A. L. Roy

School of Physics, University of Sydney 2006, Australia; and
Australia Telescope National Facility, CSIRO, PO Box 76, Epping, NSW 2121, Australia

R. P. Norris, M. J. Kesteven, E. R. Troup, J. E. Reynolds

Australia Telescope National Facility, CSIRO, PO Box 76, Epping, NSW 2121, Australia

## ABSTRACT

We have observed a sample of 157 Seyfert galaxies with a 275 km baseline radio interferometer to search for compact, high brightness temperature radio emission from the active nucleus. We obtain the surprising result that compact radio cores are much more common in Seyfert 2 than in Seyfert 1 galaxies, which at first seems to be inconsistent with orientation unification schemes. We propose a model, involving optical depth effects in the narrow-line region, which can reconcile our result with the standard unified scheme.

*Subject headings:* galaxies: active - galaxies: Seyfert - galaxies: nuclei - infrared: galaxies - radio continuum: galaxies - techniques: interferometric

## 1. INTRODUCTION

Seyfert galaxies were first separated into two distinct observational types by Khachikyan & Weedman (1971) on the basis of the width of their permitted lines, which are broader in Seyfert 1 (Sy1) than in Seyfert 2 (Sy2) galaxies. Since then a number of schemes have sought to account for the difference between these two types and, of the four schemes reviewed by Lawrence (1987), orientation unification has enjoyed particular success. This scheme, which has since been generalised by Barthel (1989), suggests that different orientations of an active galactic nucleus (AGN) relative to the line of sight can lead to very different appearances, so that the apparent differences between Sy1 and Sy2 galaxies could be due simply to the effects of our viewing angle.

This scheme invokes a dense torus of dust and molecular gas in Seyfert galaxies surrounding the nucleus at a radius between the broad-line region (BLR) and the narrow-line region (NLR). This dust obscures our view of the BLR when our line of sight lies close to the plane of the torus. In this case, only the narrow emission lines are visible since they originate from outside the torus, and the galaxy then appears as a Sy2. Light from the BLR is visible only when we view from within a cone aligned with the polar axis of the dust torus, and the galaxy then appears as a Sy1. In both cases the nucleus is the same type of object, and it is the surrounding material which causes apparent differences.

This model has received considerable support from the observations by Antonucci and Miller (1985), and Miller and Goodrich (1990), who showed that the polarized emission from some Sy2 galaxies has broad lines characteristic of a Sy1 galaxy. This is explained in terms of the torus obscuring the BLR in the Sy2, when observed directly, but the light scattered from dust or electrons above and below the torus enable us to see the BLR in scattered light.

Other schemes include the following three hypotheses:

(i) The two Seyfert types are intrinsically different. Most simply, the broad-line region gas present in Sy1 galaxies might be absent from Sy2 galaxies (e.g. Osterbrock & Koski 1976).
(ii) The Seyfert types represent evolutionary stages, in which the Sy1 nucleus switches off to produce the Sy2 state once the BLR gas has de-excited (e.g. Penston & Pérez 1984).
(iii) Sy2 narrow-line region (NLR) clouds are ionized by stellar processes (Terlevich & Melnick 1985), whereas Sy1 galaxies are ionized by a nuclear nonthermal continuum source.

Although the orientation scheme is currently favoured by many workers, the true situation is likely to be more complicated and may include elements of these other hypotheses. In this paper, we use radio observations to test the orientation model. In this model, the dust surrounding the nucleus should be optically thin at radio wavelengths, and so orientation effects should have no effect on the radio appearance. Thus we can use radio observations to check that Sy2 nuclei are intrinsically the same as Sy1 nuclei, which would test the proposition at the heart of the orientation unification model.

There are already a number of radio studies of Seyfert galaxies which compare the properties of the two Seyfert types. No significant difference was found between the radio power of Sy1 and Sy2 galaxies by Ulvestad & Wilson (1989), who used the Very Large Array (VLA) to observe a distance-limited Seyfert sample with ~2" resolution. This work improved on their earlier study of Markarian Seyferts which, they later showed, was biased against radio-weak Sy2 galaxies. Likewise, Edelson (1987) found no significant difference between the radio power of Sy1 and Sy2 galaxies in the Center for Astrophysics (CfA) Seyfert sample, also observed with the VLA. These observations seem to augur well for the unified schemes.

However, a weakness remains. The VLA observations were sensitive to extended emission from starburst activity around the nucleus in addition to the compact radio emission associated with the Seyfert core. It is possible, then, that intrinsic differences exist but are masked by stronger extended emission unrelated to the Seyfert activity. On the other hand, high-resolution, very long baseline interferometry (VLBI) observations can resolve out the extended star-forming regions, and are not sensitive to brightness temperatures less than about $10^5$ K. So, such observations are sensitive principally to nonthermal emission at $10^8$ K from the Seyfert activity, and discriminate against the $10^4$ K extended emission expected from HII regions.

We have selected 157 Seyfert galaxies using well-defined criteria, described in § 2. They were all observed with the 275 km baseline Parkes-Tidbinbilla Interferometer (PTI) at 1.7 or 2.3 GHz (Norris et al. 1988). This has resulted in a high-resolution survey with uniform sensitivity and resolution.

The PTI is sensitive to structures with brightness temperatures $>10^5$ K and sizes $<0.1"$, which corresponds to 20-200 pc over the redshift range, $0.01 < z < 0.1$, typical of the sample. Thus the PTI can detect radio emission from compact nuclear objects associated with the AGN, that have brightness temperatures ~ $10^8$ K, but is blind to the extended kpc-scale starburst regions which have typical brightness temperatures of $10^4$ K. Detectable radio supernovae are rare in starburst regions (Norris et al. 1990). Only those few which are more luminous than 400 to 40000 times Cas A would be detectable with the PTI. Whilst it may be unclear whether the nuclear emission comes from the radio 'jet' associated with the 10-pc-scale NLR clouds or from the core in individual galaxies, in either case the PTI measures radio emission associated mainly with the Seyfert activity, as opposed to that from the disc of the galaxy.

Preliminary results based on a subset of these samples have been published by Norris et al. (1992). Here we have increased the sample sizes, and refined the statistical treatment. We give fuller details of the procedure, and consider the impact of the results on the unification schemes.

We assume $H_0 = 75$ km$^{-1}$ s$^{-1}$ Mpc$^{-1}$, and $q_0 = 0.5$ throughout this paper.

## 2. SAMPLE

As our aim is to investigate whether the difference between Sy1 and Sy2 galaxies is attributable to orientation and extinction effects, it is important that our samples are not biased by these effects. Selection in the far infrared (FIR) offers the least bias of any wavelength range, since dust is probably transparent to FIR emission, unlike optical or UV wavelengths where Sy1s and Sy2s exhibit different degrees of obscuration (e.g. Spinoglio & Malkan 1989). Our primary sample is therefore FIR-selected, and is drawn from de Grijp, Miley & Lub (1987). For additional support we also used the FIR-selected sample of Norris et al. (1990).

Note however, that Burstein & Lebofsky (1986) argue that galaxies may be optically thick at 100 µm. However, this impression may be caused by a selection effect (see review by Soifer, Houck & Neugebauer 1987). Furthermore, high-resolution images of nearby spirals at 100 µm (Fitt et al. 1992) show that the 100 µm emission generally is distributed disc-wide, whereas Burstein & Lebofsky need the 100 µm emission to be dominated by the nucleus to achieve large optical depth. Bothun & Rogers (1992) found from optical and 60 µm and 100 µm radial brightness profiles, that galaxy discs are optically thin at both optical and FIR wavelengths.

For comparison, we also selected three samples on the basis of optical or mid-infrared (mid-IR) properties. We include these samples to look for any systematic differences between galaxies selected on FIR properties and those selected in the mid-IR or, more traditionally, in the optical.

The spectral classifications came from a number of sources (Norris et al. 1990; de Grijp et al. 1987; Spinoglio & Malkan 1989; Edelson 1987; Véron-Cetty & Véron 1991; Dahari & De Robertis 1988; Allen et al. 1991) and galaxies were kept for study only if all authors agreed on their classifications (except for NGC 1068 which we classed as a Sy2).

### 2.1. *FIR-Selected Samples*

#### 2.1.1. *The de Grijp et al. FIR-Selected Sample*

This sample was drawn from Table 1 of de Grijp et al. (1987), which lists galaxies with AGN-like FIR colors. Not all of these galaxies had spectroscopic classifications, and so there may be some contamination by unknown selection effects. We proceeded by assuming that the classified galaxies were representative of the whole sample. We then selected those galaxies which were classified as Sy1 or Sy2 by de Grijp et al. (1987) and which were south of +25°. Of those 128 galaxies, we rejected 14 which were inconveniently placed on the sky for observing, and of the 114 remaining another nine were rejected because various authors disagreed over their classifications, or because they were of intermediate Seyfert type. NGC 1068 was considered as a Sy2.

We rejected two objects found by the PTI to be radio-loud (3C273 at 14.4 Jy and IRAS 13451+1232 at 2.62 Jy) since they are likely to be radio-loud quasars mis-classified as Seyferts. One galaxy (IRAS 05238-4612) was rejected since the field was confused when observed with the PTI. Our analysis of the de Grijp et al. Seyferts was confined to the remaining 102 galaxies.

#### 2.1.2. *The Norris et al. FIR-Selected Sample*

For the second of the FIR-selected samples we began with the sample of Norris et al. (1990), who chose 42 FIR-cool galaxies, of which 38 were drawn from the literature (e.g. de Grijp et al. 1987), and five were drawn from the IRAS Point Source Catalog (1985) for a related project (Allen et al. 1991), with one galaxy being a member of both of these samples. All 42 galaxies satisfied the following criteria. They were: (i) optically classified as a Seyfert galaxy, (ii) detected at 60 μm and 100 μm, (iii) $L_{FIR} > 10^{10.0} L_{sun}$, (iv) redshift in the range 0.01 to 0.1, and (v) south of +25° declination.

The first group of galaxies from the literature was drawn from Table 2 of de Grijp et al. (1987), which lists catalogued AGNs detected by IRAS, with cool FIR colors. Twenty-two galaxies satisfied criteria (i) to (v) above, and had cool 25 - 60 μm colors ($S_{25 \mu m} / S_{60 \mu m} < 0.27$). Of these 22 galaxies, 11 were included by Norris et al. in their FIR-selected sample, along with II Zw 1, which was also from the de Grijp et al. Table 2 and satisfied conditions (i) to (v), above, but had $S_{25 \mu m} / S_{60 \mu m} < 0.31$. Of those 12 galaxies, we rejected one which had been given different Seyfert classifications by different authors, and we have included the remaining 11 in the present sample. We have not included the 17 galaxies drawn by Norris et al. from de Grijp et al. (1987) Table 1, since most of them (16/17) appear in our 'de Grijp et al. FIR-selected sample' (§ 2.1.1).

The next group of galaxies was drawn from Table 2 of Allen et al. (1991). Six galaxies satisfied criteria (i) to (v), above, and were located in one of two regions of sky at high galactic latitude (fields 2 and 3). Two of those six Seyferts were included by Norris et al. in their FIR-selected sample, and we have included them both in the present sample.

The final group of galaxies listed by Norris et al. (1990) were selected by them for other reasons. All nine satisfied criteria (i) to (v), above and, of those, we have included in the present sample the four which were classified unambiguously as Sy1 or Sy2.

Thus, our study of the 'Norris et al. FIR-selected sample' will be restricted to the remaining 17 selected objects. We hold a slight concern that unknown biases may have crept into this sample, as a result of the variable selection procedure. However, we have failed to detect any obvious bias, using the tests reported in § 2.4, below.

## 2.2. *Optically Selected Samples*

### 2.2.1. *The Norris et al. Optically Selected Sample*

The first of the optically selected samples was drawn from the sample selected from the literature by Norris et al. (1990). We began with their 45 galaxies which were optically classified as Seyferts and which had redshifts in the range 0.01 to 0.03. There was no attempt to make this a statistically uniform sample. Of those 45 Seyferts, 10 were rejected as being of intermediate Seyfert type or having uncertain classification, and one was rejected after observing since the PTI observation was of poor quality, leaving 34 galaxies to be considered further.

### 2.2.2. *The CfA Seyfert Sample*

The second optically selected sample was a subset of the CfA Seyfert sample. We began with the 48 galaxies in the CfA Seyfert sample listed by Edelson (1987), and kept those south of +25°. Of those 23 galaxies seven were rejected for having conflicting spectral classes, and the remaining 16 made up the sample to be considered further.

## 2.3. *The Mid-IR-Selected Seyfert Sample*

This final sample was based on the Seyfert sample compiled by Spinoglio & Malkan (1989). It was selected using the 12 μm flux density ($S_{12 \mu m} > 0.3$ Jy) since they found that "the intrinsic fraction of AGN luminosity emerging in the mid infrared is roughly constant and is hardly altered by the presence of nuclear dust". Of the 58 Seyferts in their sample, we kept the 44 galaxies south of +25°, and of those 39 were conveniently placed for observing. Nine were then rejected for having contradictory spectral classes and one radio-loud quasar (3C 273) was rejected. The remaining 29 Seyferts formed the 'mid-IR-selected sample' which we considered further.

Some sources occur in more than one sample. In such cases we have included the sources when discussing individual samples, but have deleted them from all but the earliest sample when combining levels of significance from different samples, so that the samples remain statistically independent where necessary.

The final samples are listed in Table 1.

[Table 1 appears here]

### 3. OBSERVATIONS

The samples were observed using the Parkes-Tidbinbilla Interferometer (PTI), which is described in detail by Norris et al. (1988). The interferometer consists of the 64 m antenna at Parkes with the 70 m antenna at Tidbinbilla over a 275 km baseline which provided 0.10" fringe spacing at 2290 MHz, or 0.14" at 1665 MHz.

The galaxies were observed between 1987 December and 1991 October. The details of the observations are shown in Table 2, which lists the observing dates, frequencies, polarizations, integration times, system temperatures ($T_{sys}$) and frequency standards. The quoted frequencies are the centres of the two 5-MHz bands. Most galaxies were observed once or twice at 2.3 GHz, although some had up to six cuts, or cuts at 1.7 GHz (details are given in Table 1).

[Table 2 appears here]

The data were processed using a Fourier transform fringe-search technique which produces a plot of correlated intensity as a function of fringe frequency. Flux densities were calibrated assuming that 1934-638 had a flux density of 12.2 Jy at 2290 MHz and 15.7 Jy at 1665 MHz, that Hydra A had a flux density of 27.7 Jy at 2290 MHz, and that Virgo A had a flux density of 140 Jy at 2290 MHz. These values were derived from the Baars et al. (1977) measurements of Cas A, and used the comparison of 1934-638 to Hydra A by Reynolds (1992).

A system temperature ($T_{sys}$) correction was applied to the flux densities to correct for the nonlinearity which occurs when the source is strong enough to raise $T_{sys}$ significantly. The fringe visibility measured by the PTI ($S_{PTI}$) depends not only on the correlated source flux density (S) but also on $T_{sys}$, with the relationship:

$$S_{PTI} = \frac{kS}{\sqrt{(S + T_{sys_1})(S + T_{sys_2})}} \qquad (1)$$

where:     $S_{PTI}$   is dimensionless,
              S      is in Jy,
              $T_{sys}$   is in Jy,
              k      is a dimensionless scaling constant.

Note that equation (1), which applies to all 1-bit sampling interferometers, reduces to the $T_{sys}$ correction given by Cohen et al. (1975) in the limit of small S. In practice, the departure from linearity becomes significant only for the stronger calibrators (>~5 Jy). The calibration constant, k, was determined from the calibrator observations using equation (1), and was then used to calibrate the program source observations, assuming $S \ll T_{sys}$. Nominal values of $T_{sys}$, based on other observations, were used. The flux density calibration was insensitive to the adopted value of $T_{sys}$, and errors in the values, or changes due to zenith angle, will result in an error of $<\sim 3\%$.

Finally, the data were corrected for Rice bias, which would otherwise have caused the flux densities to be overestimated slightly at low signal-to-noise ratios. The overestimate occurs because Gaussian noise in the real and imaginary components of the fringe visibility is non-Gaussian when represented as amplitude and phase, and amounts to 2% at the detection limit.

The delay and phase beams had FWHM of 55" and 13' along the minor axes, which were larger than the uncertainty in the IRAS galaxy positions from the IRAS Point Source Catalog. A 1-$\sigma$ position error (~ 14") should cause an average decorrelation of 4%.

To combine flux densities from different observations, we first converted any 1.7-GHz flux densities or limits to 2.3 GHz assuming a spectral index of -0.7, and then quoted the average of the detections, with all detections weighted equally. Where there were upper limits only, we quoted the lowest upper limit.

The resulting flux densities are estimated to have uncertainties of +6%-7% rms random multiplicative, with 0.6 mJy rms random additive, and a systematic multiplicative uncertainty of 11% rms due mainly to uncertainty in the Baars et al. flux density scale.

Each galaxy was typically observed for 10 or 20 min coherently, with two 5-MHz bands. This produced a 5-$\sigma$ sensitivity limit of 8 mJy at 1665 MHz and 3 mJy at 2290 MHz. We tested for coherence loss and found that it was insignificant (< 5%), and therefore did not apply coherence corrections.

## 4. RESULTS

### 4.1. *Properties of the Samples*

We found no significant difference between the disc radio luminosities of Sy1s and Sy2s in the five samples combined. The disc emission at 2.4 GHz was measured for 97 of the 157 galaxies by using the compact array of the Australia Telescope, in a related project (Roy et al. 1994), and by subtracting off the core flux density measured by the PTI. We constructed radio luminosity functions for the two Seyfert types, and compared them using survival analysis two-sample tests, with ASURV rev 1.2 (La Valley, Isobe & Feigelson 1992), which implements the methods presented in Feigelson & Nelson (1985). The tests found that any differences had a level of significance between only 14% (logrank test) and 30% (Gehan test). Since the level is not less than the critical value of 5%, the distributions of integrated disc radio luminosity of the two Seyfert types are indistinguishable.

The redshift distributions also showed no significant difference between Sy1s and Sy2s for the FIR sample. To make the comparison, we overlayed plots of the Kaplan-Meier estimator[1] for Sy1s and for Sy2s (Fig. 1a) and found agreement within the 95% confidence intervals. The difference between the redshift distributions was significant at between the 34% (Gehan test) and 45% (logrank test) levels of significance.

---

[1] The Kaplan-Meier estimator provides an efficient, non-parametric estimate of the cumulative distribution function, for data which may include upper limits as well as detections. It treats objects with limits by assuming that they are distributed in the same way as the detected objects, and that the censoring is random.

On the other hand, in the optically selected samples, Sy1s were more distant than Sy2s (Fig. 1b), with a significance between the 0.3% (Peto-Peto test) and 2% (logrank test) levels. This effect is well known in optical-flux-limited Seyfert samples (e.g. Dahari & De Robertis 1988) and it arises since Sy2s are less luminous optically than Sy1s. This bias makes it difficult to compare properties of the Seyfert types for optical-flux-limited samples and so, to avoid ambiguous interpretation, we will work mainly with the FIR-selected sample which does not suffer from this effect.

[Figure 1 appears here]

We compared the FIR luminosity functions of Sy1s to Sy2s for the FIR-selected and mid-IR-selected samples and the optically selected samples, and found that any differences had a level of significance of between 24% (Peto-Prentice test, on the mid-IR-selected sample) and 90% (logrank test, on the FIR-selected sample; i.e. up to 90% of random samples would display a bigger difference than that observed). We therefore conclude that the two Seyfert types were indistinguishable in FIR luminosity in all the samples.

Finally, we compared the [OIII] luminosities of the Sy1s to the Sy2s in the de Grijp et al. FIR-selected sample, for which [OIII] data were readily available. We overlayed plots of the Kaplan-Meier estimator for the [OIII] luminosities without reddening correction, and found that the distributions were similar, but there may be a tendency for Sy2s to be lower luminosity than Sy1s (Fig. 2). Formally, the difference had a level of significance between the 2.3% (logrank test) and the 12% (Gehan test) levels. This marginally significant result may be consistent with that of Dahari and De Robertis (1988), who found that optically selected Sy1s and Sy2s had indistinguishable [OIII] luminosity distributions before reddening correction. We expect that Sy1s and Sy2s in our optically and mid-IR-selected samples are similar to those of Dahari and De Robertis, and so should also have similar [OIII] luminosity distributions.

[Figure 2 appears here]

In summary, all the galaxies in these five Seyfert samples are classified clearly as Sy1 or Sy2, with none having intermediate spectral class. The two Seyfert types have essentially the same redshift distributions in the FIR-selected sample, and the radio luminosity functions and FIR luminosity functions are indistinguishable between types 1 and 2 in all the samples.

### 4.2. *The Detection Rates of Compact Cores*

All five Seyfert samples were observed with the PTI and the detection rates are illustrated by Figure 3, which shows the Sy1 and Sy2 detection rates separately for the FIR-selected sample and for the optical+mid-IR-selected sample.

In both the FIR- and the optical+mid-IR-selected samples almost half of the Sy2s were detected, whereas only one quarter of the Sy1s were detected. We used the 'difference-of-two-proportions' test (Appendix § a) with the Yates continuity correction (Appendix § b) to quantify this apparently significant result. The null hypothesis $H_0$:Sy1=Sy2 was rejected in favour of the alternative hypothesis $H_1$:Sy1≠Sy2 at the 2.0% level of significance for the FIR-selected sample, but only at the 14.7% level for the optically selected sample. So, the FIR-selected sample displays a significant difference between the Sy1 and Sy2 detection rates, consistent with the appearance of Figure 3.

The result was weak for the optical+mid-IR-selected sample because the sample size was relatively small (64). Had the FIR-selected sample been as small as the optically selected sample and the detection rate remained the same, the significance would have been 10%.  Thus we do not consider this weak result, on its own, to be evidence for different behaviour between the FIR-selected and optical+mid-IR-selected samples, but merely an artefact of the different sample sizes.

[Figure 3 appears here]

The result for the optically selected sample requires a caveat: the type 2 Seyferts are closer, on average, than the type 1s.  The effect is in the correct sense, and is about the right size, to explain the difference in the core detection rates in the optically selected sample.  However, the detection rates in the FIR-selected sample cannot be explained so simply, since the redshift distributions of the two Seyfert types are much too similar (Fig. 1).

To examine further this curious difference between FIR-selected Sy1s and Sy2s, we looked for the effect in the two FIR-selected samples individually.  We divided the FIR-selected sample from Figure 3 into the de Grijp et al. and the Norris et al. FIR-selected samples, and compared the detection rates of Sy1s and Sy2s in the two samples.  The results are illustrated in Figure 4.

In both the de Grijp et al. and the Norris et al. FIR-selected samples, almost half of the Sy2s were detected, as opposed to only one quarter or less of the Sy1s.  To estimate the significance of this result, we strengthened the 'difference-of-two-proportions' test in three ways.  Firstly, we considered the two FIR-selected samples in Figure 4 individually, rather than grouping them together.  This led to a stronger test since the result could be seen in both samples separately.  Secondly, we applied a one-tailed test to the Norris et al. FIR-selected sample.  The method and advantage are explained in Appendix § c.  Lastly, we combined the results of two FIR-selected samples to produce one level of significance (Appendix § d).  These improvements strengthened the result and better reflected the confidence which came from seeing the result in independent samples.  We used these improvements to combine the two FIR-selected samples as described in § e of the Appendix, and we obtained a significance level of 1.2%.  The observed difference between the Seyfert types is very unlikely to occur by chance.

[Figure 4 appears here]

In summary, Sy1s and Sy2s from the FIR-selected Seyfert sample have a significantly different detection rate of compact, high brightness temperature radio structure.  The combined optical and mid-IR-selected sample may also show a lack of cores in Sy1s, but that effect may be caused simply by the fact that optically selected Sy2s are closer on average than optically selected Sy1s.  We tested other ways of separating the galaxies into the two Seyfert classes and found no significant change to our result.

5. DISCUSSION

5.1. *The Challenge to Models of Seyfert Galaxies*

We have found, surprisingly, that FIR-selected Sy1s were detected significantly less frequently by the PTI than were Sy2s.  This result is inconsistent with the standard orientation model, which has considerable evidence in its support (e.g. Miller & Goodrich 1990) and which would predict an equal fraction of detections for Sy1s and for Sy2s.  It is also in the opposite sense expected from alternative models in which Sy1s are expected to have more energetic cores than Sy2s, or may be relativistically beamed.  We are left in the unsatisfactory position where there is no existing model which is consistent with our observation.

### 5.2. *A Model to Explain our Observations*

Here we offer one model which attempts to reconcile our result with the standard model of orientation unification. This model is based on that first proposed by Norris et al. (1992), and invokes the optical depth of the NLR clouds at 2.3 GHz, due to free-free absorption, to reconcile these results with the standard unified model discussed above. However, there are two distinct mechanisms which may contribute to the resulting radio appearance, and we now discuss these separately.

Mechanism 1: Obscuration by the narrow-line region

In Sy1 galaxies we view the radio core through the full depth of the NLR since our view passes nearly down the NLR axis. If the optical depths are above unity, as discussed below, then the NLR clouds will block our view of the core and of each other (Fig. 5a).

[Figure 5 appears here]

In Sy2 galaxies we view the compact core through very little of the NLR because our line of sight is nearly perpendicular to the NLR axis. Since there are fewer NLR clouds which could block the view of either the core or other clouds, it follows naturally that the PTI should detect nuclear emission more often in Sy2s than in Sy1s. This argument holds even if the radio core comes from an inner NLR cloud rather than the AGN, although it becomes weaker if the radio core is located in an outer NLR cloud.

Mechanism 2: Obscuration by individual NLR clouds

This model also invokes free-free absorption by the NLR clouds but with a different geometry from that used by our first model, and applies only if the PTI cores are located near the NLR clouds rather than in the BLR. It is illustrated by Figure 5b. It has been noted by several authors that compact radio emission is closely associated with individual NLR clouds, and that in one case there may be a small offset between the radio and NLR emission (Ulvestad, Wilson & Sramek 1981; Whittle et al. 1986; Whittle et al. 1988; Haniff, Wilson & Ward 1988; Evans et al. 1991). When observing Sy1s, radio emission from components on the far side of the nucleus would be blocked from view by the NLR clouds with which they are associated if the NLR clouds lie closer to the nucleus than do the radio components, and so only the components on the near side would be visible in the radio. Sy2s, on the other hand, are viewed perpendicular to the axis of the NLR, and so our view of the radio components would not be obscured by the NLR clouds. If the PTI cores are located within these radio components, then the radio power of Sy1 cores would appear to be half that of Sy2 cores and the PTI detection rate would be lower for Sy1s than for Sy2s. A similar mechanism would operate if the radio knots were on the core side of the NLR clouds.

Which of these two mechanisms will dominate will depend on the location of the PTI cores (in the BLR or in the NLR), the opening angle of the cone (discussed below), and the filling factor. Observations do not yet permit us to say which of these two related mechanisms will dominate, and it may even vary from source to source.

We note that both mechanisms in this model predict that the NLR clouds will become optically thin at higher frequency, and so the model can be tested by repeating the experiment at a sufficiently high frequency and checking that there is then no difference between the Sy1 and Sy2 core detection rates.

We have attempted to anticipate potential criticisms, and arguments to support our model are given in the following sections.

### 5.3. *The Nature of the Radio Emission from Seyfert Galaxies*

There is now strong evidence that much of the radio emission from Seyfert galaxies is closely associated with the NLR clouds, although the exact mechanism is still a matter of debate. For example, Evans et al. (1991) have compared the VLA image of NGC 1068 with a Hubble Space Telescope (HST) image made in the light of [OIII] and find that knots of radio emission are coincident with NLR clouds. This confirms earlier work (e.g. Wilson & Willis 1980; Whittle et al. 1986; Haniff et al. 1988; Whittle et al. 1988 and references therein; Veilleux 1991) which indicate a close link between NLR clouds and radio emission. Models discussed by Veilleux (1991) and by Pedlar, Dyson & Unger (1985) and references therein, attempt to explain the excitation and the close relationship between the NLR clouds and radio emission.

Despite this strong evidence that much of the radio emission originates from the NLR, it is still not clear where the compact radio cores, detected by VLBI and PTI observations, are located. In some cases (e.g. Ulvestad et al. 1981) the compact radio core is coincident with the optical nucleus to about 0.3" and, in these cases, the radio emission may be either from the BLR or from the AGN itself, presumably as a low-luminosity analog to the cores of quasars and radio galaxies. However, 0.3" corresponds to 60-600 pc at redshifts of 0.01 to 0.1, and so the case for true alignment is not strong. In other cases, compact VLBI cores are seen in the inner part of the NLR, usually as distinct multiple sources (e.g. Whittle et al. 1986; Kukula et al. 1993). In these cases it appears that the NLR clouds close to the BLR contain unresolved hot-spots of radio emission. NGC 1068 (Evans et al. 1991) appears to belong to this latter category. However, most VLBI observations, like those presented here, do not have sufficient information to locate the radio core relative to the optical nucleus, and we regard the question as open. In subsequent discussion, we will refer simply to the radio "core", although this may be located either in the BLR or in the inner part of the NLR. The precise location does not affect our argument significantly.

### 5.4. *Arguments to Support the Simple Model*

#### 5.4.1. *The optical depth of the NLR*

The electron temperature in the NLR clouds is estimated from [OIII] line ratios (Osterbrock 1987) to be between $10^4$ and $5 \times 10^4$ K, and the electron density is variously estimated to be $10^{9.5 \pm 1}$ m$^{-3}$ (Osterbrock 1987), or between $10^8$ and $10^{13}$ m$^{-3}$ (Lawrence 1987; Haniff et al. 1988). The size of the clouds is estimated to be between 6 and 12 pc in NGC 1068 (Evans et al. 1991), and the size of the NLR is estimated to be between 50 pc (Evans et al.) and 1 kpc with a filling factor ~0.1 (Lawrence 1987). These large uncertainties naturally lead to a large uncertainty in the optical depth, but since they formally give a free-free optical depth per cloud between 0.01 and $10^3$ at 2.3 GHz, using Osterbrock's electron densities (which is consistent with the optical depth of 16 derived by Whittle et al. 1986), we have some justification in assuming the clouds to be optically thick. The filling factor of 0.1 implies a covering factor (depending on the size of the clouds) approaching unity when the core is viewed through the full depth of the cone.

In the most extreme case, in which the optical depth is $10^3$ at 2.3 GHz, the clouds become optically thin ($\tau < 0.1$) to free-free absorption above a critical frequency of 200 GHz. However, the critical frequency will be very much lower than this for less extreme NLR conditions, and will even be less than 2.3 GHz for the smallest values of density and cloud size.

We note in passing that a similar calculation for the BLR clouds, assuming $T_e \leq 4 \times 10^4$ K, $N_e \sim 10^{14}$-$10^{18}$ m$^{-3}$, r~0.003 pc, filling factor ~0.01, (Lawrence 1987; Osterbrock 1987; Zheng 1992) gives a corresponding optical depth for the BLR clouds of $10^8$ to $10^{11}$, and so we can be confident that the BLR clouds are optically thick at 2.3 GHz. Despite the large optical depth, radio emission can still be detected from the nuclei of Seyfert galaxies since the corresponding covering factor of BLR clouds is much less than unity (e.g. 5%-30% is the covering factor estimated by Netzer & Laor 1993).

### 5.4.2. *The optical depth of the torus*

The inner edge of the dust torus can, in principle, be ionized by the nucleus, and Krolik & Lepp (1989) have suggested that this will be optically thick at 2.3 GHz. However, this depends on quantities such as ionization fraction which are even more poorly known for the torus than for the NLR, and so there is considerable uncertainty in this result. Taking the range of values adopted by Krolik and Lepp in their paper gives an optical depth ranging from 0.02 to 28, and only for their most extreme models does the torus become optically thick. Thus, whilst admitting the possibility that in extreme cases the torus could become optically thick, there is no reason to assume that this will be common. Furthermore, their result, which predicts that we should not see the AGN in Sy2 galaxies, is inconsistent with the observations cited above in which VLBI cores are observed more frequently in Sy2 than in Sy1 galaxies.

### 5.4.3. *The optical depth of NLR clouds*

Seyfert galaxy cores generally have steep spectral indices (e.g. Unger et al. 1986), whereas absorbed spectra are flat. This has been used to argue that our line of sight to the nuclear radio sources is not obscured by NLR clouds (Wilson 1993). However, the clouds may still be optically thick since the strongly absorbed flat spectrum component would also be very much weaker than the steep spectrum component directly visible from radio knots on the near side of the core. Then one would see only the steep spectrum cores.

X-ray spectra of Sy1 nuclei sometimes diagnose absorbing columns in the NLR which are smaller than those we need to block radio emission (e.g. Pounds et al. 1990; Rao, Singh & Vahia 1992). However, an absorbing column large enough to block radio emission has been seen in at least one Seyfert 1 (Turner et al. 1993). Furthermore, in many cases the NLR clouds will not intersect the line of sight to the X-ray source since their covering factor is less than unity. Thus X-rays will not always suffer the large absorption expected from NLR clouds. In contrast, in our second model the radio knots are closely associated with the NLR clouds, and so the covering factor seen by the radio knot is very much larger than that seen by the X-ray source. Thus we are not deeply concerned by the low X-ray absorbing columns.

### 5.4.4. *The NLR filling factor*

The filling factor of the NLR is probably very small (Lawrence 1987; Evans 1991). Assuming a filling factor of 0.1 implies a covering factor of ~0.5, and so not all Sy1 cores will be obscured by the NLR. Such large covering factors are in conflict with the diagnosis from photoionization models, which estimate the covering factor to be closer to ~0.01. However, Netzer & Laor (1993) show that dust in the NLR can strongly suppress line emission, which causes these models to underestimate the covering factor. The true factor is more nearly 0.2 to 0.5. Thus, although some Sy1 cores should be observable in the radio, the fraction of those detected should be less than that for Sy2 galaxies. This is consistent with our observations.

### 5.4.5. *The opening angle of the NLR cone*

The opening angle of the NLR cone has been variously estimated to be 65° (Evans et al. 1991), 86° (Krolik & Begelman 1988) and 120° (Krolik & Begelman 1986; Bregman 1990). In cones with such large opening angles, a compact core placed at an arbitrary position within the cone will have less extinction along the cone axis than it will when viewed perpendicular to the axis. (Although this is clearly not true for clouds at the cone apex, most clouds lie nearer the cone base where the path lengths are larger in the direction perpendicular to the cone axis.) Therefore, for our mechanism (1) to operate, we require that the compact cores are located at or close to the apex of the cone and that the torus is optically thin and the NLR is optically thick at 2.3 GHz. At present this is consistent with observations, but mechanism (1) of our model would cease to be important if the cores were found to be located a large distance from the AGN.

## 6. CONCLUSION

We have observed FIR-selected, mid-IR-selected and optically selected samples of Seyfert galaxies with a long-baseline interferometer. We have demonstrated that compact radio structures are much more common in Seyfert 2 than in Seyfert 1 galaxies in the FIR-selected samples, and possibly in the combined mid-IR- and optically selected sample as well.

We have proposed a model which explains our observations in the framework of the orientation unification model, after considering the radio optical depth of the NLR clouds which surround the radio-emitting regions of the core and NLR.

This scenario makes the testable prediction that the NLR clouds might become optically thin at higher frequencies, and so further observations at a sufficiently high frequency should find no difference between the Sy1 and Sy2 detection rates.

## ACKNOWLEDGMENTS

We thank David Allen who has been involved in this project since its inception, for the many hours he spent with us at the telescope, for his insight during discussions, and for his humour at early hours of the morning.

We thank the staff and Director of the Tidbinbilla Tracking Station for their assistance with the observations presented here.

This research has made use of the NASA/IPAC Extragalactic Database (NED) which is operated by the Jet Propulsion Laboratory, Caltech, under contract with the National Aeronautics and Space Administration.

## APPENDIX

This appendix justifies the use of the 'difference-of-two-proportions' test, explains how it was applied, the continuity correction, the use of one- and two-tailed tests, and how different samples were combined to strengthen the test.

A.1. *The 'Difference-of-Two-Proportions' Test*

We considered the observation of a Seyfert galaxy by the PTI as a binomial experiment with the outcomes being either a detection or an upper limit. The probability of detection (p) is an intrinsic property of the parent population from which the galaxy was selected, and after drawing a sample of size n, the probability of observing a number (m) of detections is given by the binomial distribution. The binomial distribution is well approximated by the normal distribution if the sample size is sufficiently large. More precisely, the approximation is good when both np and n(1-p) are greater than ~ five (e.g. Glantz 1992, p120), which was always the case for the samples we considered (we grouped samples where necessary to ensure this condition was met). We could then use a range of tests developed for normally distributed random variables.

We used the 'difference-of-two-proportions' test to compare two populations to determine whether they differed with respect to the detection rate. We tested the null hypothesis $H_0:p_1=p_2$, where $p_1$ and $p_2$ are the observed detection rates of the Sy1 and Sy2 samples. Since both $p_1$ and $p_2$ have approximately normal distributions and the null hypothesis assumed that the distributions have equal means, then the distribution of $p_1 - p_2$ is also approximately normal and has zero mean. In general, the observed value of $p_1 - p_2$ will differ from zero, and the difference, in standard deviations, gives a measure of how often such a difference would occur at random. The test estimates the standard deviation of the distribution of $p_1 - p_2$ from the sample sizes and detection rates. The derivation of the test statistic is given in elementary statistics texts (e.g. Glantz 1992, pp121-124), which yield the result:

$$z = \frac{p_1 - p_2}{\sqrt{p(1-p)(\frac{1}{n_1} + \frac{1}{n_2})}} \qquad (A1)$$

where  $z$ = the test statistic
  $p_1$ = observed detection rate of Sy1,
  $p_2$ = observed detection rate of Sy2,
  $n_1$ = Sy1 sample size,
  $n_2$ = Sy2 sample size,
  $p$ = the best estimate of the intrinsic detection rate,
    = (# Sy1 detections + # Sy2 detections) / ($n_1 + n_2$).

The standard normal table is used to determine the probability with which z would exceed the value calculated using equation (A1) if the two samples had been drawn at random from the same parent population. This result is the level of significance for the test.

A.2. The Yates Correction for Continuity

The test statistic (z) was referred to the standard normal tables, assuming it was a continuous variable. In fact, it takes on discrete values, and the approximation made the result seem more significant than it actually was, often by as much as a factor of two. The 'Yates correction for continuity' compensates for this error. The test statistic for the 'difference-of-two-proportions' test with Yates correction is given by Glantz (1992) p124, for example, as:

$$z = \frac{|p_1 - p_2| - \frac{1}{2}(\frac{1}{n_1} + \frac{1}{n_2})}{\sqrt{p(1-p)(\frac{1}{n_1} + \frac{1}{n_2})}} \quad (A2)$$

where the notation is given in equation (A1).

This correction improves the estimate provided $0.2 < p < 0.8$, and is used throughout this paper.

### A.3. *One-tailed vs Two-tailed Tests*

The first time we compare two Seyfert samples for a difference between their core detection rates, there are three possible outcomes (Sy1 = Sy2, Sy1 < Sy2, and Sy1 > Sy2) because we have no prior information about the core properties of Seyferts. In this case we test the null hypothesis $H_0$:Sy1=Sy2 against the alternative hypothesis $H_1$:Sy1≠Sy2. This is a two-tailed test.

Once we have seen, say, a lack of cores in Sy1s in the first sample we test, and subsequent samples are being used to confirm the result, then the question changes from: "Is there a difference?" to: "Are Sy1s detected less frequently in this sample too?" The alternative hypothesis becomes $H_1$:Sy1<Sy2 (a one-tailed test), and the null hypothesis remains unchanged, $H_0$:Sy1=Sy2 (e.g. Hoel 1966).

To apply a one-tailed test, one takes the critical region to be the area under only one tail of the normal distribution lying beyond the test statistic, instead of the area under both tails as used in the two-tailed test. The one-tailed test is stronger than the two-tailed test.

Two-tailed tests are used unless otherwise stated.

### A.4. *Combining Levels of Significance*

We combined the levels of significance from the different samples by multiplying the individual levels of significance. The samples must be independent before they can be combined in this way, and so we ensured independence by discarding from later samples any galaxies which had already been counted towards the earlier samples.

### A.5. *Combining the samples*

The de Grijp et al. FIR-selected sample showed that Sy1s were detected less frequently than were Sy2s, significant at the 5.1% level. Therefore, we could test this specific hypothesis on the Norris et al. FIR sample. We therefore used a one-tailed test on the Norris et al. FIR-selected sample (after having omitted the one galaxy which was also in the de Grijp et al. sample) to test whether it also showed a lack of Sy1 cores, and the result was significant at the 22.8% level. Finally we multiplied together these two levels of significance and found that Sy1 cores were detected less frequently than were Sy2 cores, significant at the 1.2% level.

15TABLE 1
Seyfert Galaxies Observed with the PTI

| IRAS name (1) | alias (2) | redshift (3) | class (4) | sample (5) | PTI flux / Jy (6) | no. of cuts (7) | (8) |
|---|---|---|---|---|---|---|---|
| 00037+1955 | MRK 335 | 0.0258 | Sy1 | EN1S | 0.005 | 2 | 3 |
| 00198-7926 |  | 0.0728 | Sy2 | DS | 0.004 | 0 | 1 |
| 00321-0018 |  | 0.0420 | Sy2 | D | 0.014 | 0 | 1 |
| 00333-8156 |  | 0.1271 | Sy2 | D | < 0.002 | 0 | 1 |
| 00350+0000 | MRK 955 | 0.0348 | Sy2 | N1 | < 0.005 | 0 | 1 |
| 00392-7930 | ESO 12- G 21 | 0.0328 | Sy1 | N2 | < 0.003 | 0 | 1 |
| 00447+1425* | MRK 1146 | 0.0389 | Sy1 | N1 | < 0.005 | 0 | 1 |
| 00492+1709* | MRK 1148 | 0.0640 | Sy1 | N1 | < 0.005 | 0 | 1 |
| 00509+1225 | I ZW 1 | 0.0610 | Sy1 | DEN1S | 0.004 | 1 | 2 |
| 00521-7054 |  | 0.0688 | Sy2 | D | 0.013 | 0 | 1 |
| 00598-1956 | ESO 541-IG 12 | 0.0560 | Sy2 | D | < 0.004 | 0 | 1 |
| 01072-0348 |  | 0.0546 | Sy2 | D | 0.007 | 0 | 3 |
| 01112+1300 | MRK 975 | 0.0491 | Sy1 | N1 | < 0.007 | 0 | 1 |
| 01113-1506* | MRK 1152 | 0.0522 | Sy1 | N1 | < 0.010 | 0 | 1 |
| 01194-0118 | II ZW 1 | 0.054 | Sy1 | N2 | < 0.003 | 0 | 1 |
| 01248+1855 | MRK 359 | 0.0167 | Sy1 | DN1 | < 0.005 | 0 | 1 |
| 01346-0924 | MCG -2- 5- 22 | 0.0698 | Sy2 | D | 0.011 | 0 | 1 |
| 01356-1307 |  | 0.0404 | Sy2 | D | 0.030 | 0 | 1 |
| 01378-2230 |  | 0.0861 | Sy1 | D | < 0.004 | 0 | 1 |
| 01413+0205 | MRK 573 | 0.0172 | Sy2 | DEN1 | < 0.005 | 2 | 1 |
| 01475-0740 |  | 0.0177 | Sy2 | D | 0.256 | 0 | 1 |
| 01572+0009 | MRK 1014 | 0.1628 | Sy1 | D | 0.008 | 0 | 2 |
| 02043-5525 | ESO 153- G 20 | 0.0198 | Sy2 | D | < 0.004 | 0 | 1 |
| 02090-4956 | FAIRALL 377 | 0.0475 | Sy2 | D | 0.006 | 0 | 1 |
| 02120-0059 | NGC 863 | 0.0263 | Sy1 | EN1 | 0.005 | 1 | 1 |
| 02304+0012 | UGC 2024 | 0.0221 | Sy2 | D | < 0.004 | 0 | 1 |
| 02321-0900 | NGC 985A/B | 0.0431 | Sy1 | D | < 0.003 | 0 | 2 |
| 02366-3101 | ESO 416- G 5 | 0.0620 | Sy1 | D | 0.006 | 0 | 1 |
| 02389+0658* | MRK 595 | 0.0275 | Sy1 | N1 | < 0.005 | 0 | 1 |
| 02401-0013 | NGC 1068 | 0.0038 | Sy2 | DES | 0.086 | 1 | 1 |
| 02526-0023 | NGC 1143/4 | 0.0282 | Sy2 | EN1S | 0.004 | 1 | 1 |
| 02537-1641 |  | 0.0315 | Sy2 | D | < 0.003 | 0 | 1 |
| 02580-1136 | MCG -2- 8- 39 | 0.0296 | Sy2 | D | < 0.005 | 0 | 1 |
| 03059-2309 | NGC 1229 | 0.0355 | Sy2 | D | < 0.003 | 0 | 1 |
| 03106-0254 |  | 0.0272 | Sy2 | D | 0.011 | 0 | 1 |
| 03109-5131 | ESO 199-IG 23 | 0.0778 | Sy2 | D | 0.005 | 0 | 1 |
| 03125+0119 | KUG 312+ 13 | 0.0233 | Sy2 | D | 0.005 | 0 | 1 |
| 03202-5150 | FAIRALL 299 | 0.0578 | Sy2 | D | 0.010 | 0 | 1 |
| 03222-0313 | MRK 607 | 0.0087 | Sy2 | DS | < 0.004 | 0 | 1 |
| 03229-0618 | MRK 609 | 0.0341 | Sy1 | N2 | < 0.003 | 0 | 1 |
| 03230-5800 |  | 0.0437 | Sy2 | D | < 0.004 | 0 | 1 |
| 03238-6054 | ESO 116- G 18 | 0.0185 | Sy2 | D | < 0.004 | 0 | 1 |
| 03317-3618 | NGC 1365 | 0.0055 | Sy1 | S | 0.004 | 1 | 1 |
| 03335-5625 |  | 0.0785 | Sy2 | D | < 0.004 | 0 | 1 |
| 03348-3609 | NGC 1386 | 0.0031 | Sy2 | S | 0.004 | 1 | 1 |
| 03355+0104 |  | 0.0396 | Sy2 | D | 0.012 | 0 | 1 |
| 03362-1641 |  | 0.0369 | Sy2 | D | < 0.003 | 0 | 1 |
| 03380-7113 | ESO 54-IG 15 | 0.0475 | Sy2 | D | < 0.004 | 0 | 1 |
| 04124-0803 |  | 0.0379 | Sy1 | D | < 0.005 | 0 | 1 |
| 04189-5503 | NGC 1566 | 0.0050 | Sy1 | S | 0.005 | 1 | 2 |
| 04229-2528 |  | 0.0436 | Sy2 | D | < 0.003 | 0 | 1 |
| 04305+0514 | 3C120 | 0.0327 | Sy1 | DS | 0.244 | 0 | 1 |
| 04339-1028 | MRK 618 | 0.0344 | Sy1 | DN1S | < 0.003 | 1 | 1 |
| 04385-0828 |  | 0.0149 | Sy2 | DS | 0.006 | 0 | 1 |
| 04392-5946 | ESO 118-IG 33 | 0.0577 | Sy2 | D | < 0.004 | 0 | 1 |
| 04448-0513 |  | 0.0442 | Sy1 | D | < 0.004 | 0 | 1 |
| 04461-0624 | NGC 1667 | 0.0153 | Sy2 | SN2 | < 0.003 | 0 | 2 |

TABLE 1-*Continued*

| IRAS name (1) | alias (2) | redshift (3) | class (4) | sample (5) | PTI flux / Jy (6) | no. of cuts (7) | (8) |
|---|---|---|---|---|---|---|---|
| 04493-6441 |  | 0.0600 | Sy1 | D | < 0.005 | 0 | 1 |
| 04502-0317 |  | 0.0158 | Sy2 | D | < 0.004 | 0 | 1 |
| 04505-2958 |  | 0.2860 | Sy1 | D | < 0.002 | 0 | 2 |
| 04507+0358 | CGCG 420- 15 | 0.0297 | Sy2 | D | < 0.004 | 0 | 1 |
| 04575-7537 | ESO 33- G 2 | 0.0181 | Sy2 | D | < 0.005 | 0 | 1 |
| 05136-0012 | UGC 3271 | 0.0312 | Sy1 | D | < 0.004 | 0 | 2 |
| 05177-3242 | ESO 362- G 18 | 0.0126 | Sy1 | D | < 0.004 | 0 | 1 |
| 05189-2524 |  | 0.0415 | Sy2 | S | < 0.003 | 1 | 1 |
| 05218-1212 |  | 0.0490 | Sy1 | DN1 | < 0.004 | 1 | 1 |
| 05497-0728 | NGC 2110 | 0.0160 | Sy2 | N1 | 0.035 | 1 | 1 |
| 05595-5756 |  | 0.0383 | Sy2 | D | < 0.004 | 0 | 2 |
| 06115-3240 |  | 0.0500 | Sy2 | DN1 | < 0.003 | 1 | 1 |
| 06317-6403 |  | 0.0485 | Sy2 | D | < 0.004 | 0 | 1 |
| 06563-6529 | FAIRALL 265 | 0.0295 | Sy1 | D | < 0.003 | 0 | 1 |
| 08255-7737 | ESO 18- G 9 | 0.0175 | Sy2 | D | < 0.003 | 0 | 1 |
| 08277-0242 |  | 0.0404 | Sy2 | D | 0.006 | 0 | 1 |
| 08518+1752 | MRK 1220 | 0.0640 | Sy1 | D | < 0.004 | 0 | 1 |
| 09143+0939 |  | 0.047 | Sy1 | N2 | < 0.004 | 0 | 1 |
| 09182-0750 | MCG -1-24- 11 | 0.0198 | Sy2 | D | 0.015 | 0 | 1 |
| 09233+1256 |  | 0.0285 | Sy1 | N1 | < 0.004 | 0 | 2 |
| 09305-8408 |  | 0.0628 | Sy2 | D | 0.012 | 0 | 1 |
| 09344+0119* | MRK 707 | 0.0498 | Sy1 | N1 | < 0.006 | 0 | 1 |
| 09379+2127* | MRK 403 | 0.0241 | Sy2 | N1 | < 0.006 | 0 | 1 |
| 09432-1405 | NGC 2992 | 0.0077 | Sy1 | S | 0.006 | 1 | 1 |
| 09435-1307 |  | 0.1310 | Sy2 | D | < 0.004 | 0 | 1 |
| 09497-0122 | MRK 1239 | 0.0194 | Sy1 | DN1S | 0.028 | 0 | 1 |
| 09572+1317* | MRK 1243 | 0.0353 | Sy1 | E | < 0.003 | 0 | 1 |
| 10295-3435 | NGC 3281 | 0.0110 | Sy2 | N2 | 0.023 | 0 | 1 |
| 10422+0651* | NGC 3362 | 0.0227 | Sy2 | E | < 0.005 | 0 | 1 |
| 10459-2453 | NGC 3393 | 0.0137 | Sy2 | D | 0.016 | 0 | 1 |
| 10511-2723 |  | 0.1599 | Sy2 | D | < 0.004 | 0 | 1 |
| 11058-1131 |  | 0.0546 | Sy2 | D | < 0.004 | 0 | 1 |
| 11083-2813 | ESO 438- G 9 | 0.0245 | Sy1 | N2 | < 0.003 | 0 | 1 |
| 11215-2806 |  | 0.0135 | Sy2 | D | < 0.003 | 0 | 1 |
| 11249-2859 | MCG -5-27- 13 | 0.0234 | Sy2 | D | < 0.003 | 0 | 1 |
| 11365-3727 | NGC 3783 | 0.0107 | Sy1 | DN1 | < 0.005 | 1 | 1 |
| 11581-2033 |  | 0.0621 | Sy1 | N2 | 0.004 | 0 | 2 |
| 12146+0728* | NGC 4235 | 0.0077 | Sy1 | E | < 0.006 | 1 | 0 |
| 12370-0504 | NGC 4593 | 0.0078 | Sy1 | DS | < 0.005 | 1 | 1 |
| 12381-3628 | IC 3639 | 0.0125 | Sy2 | DS | 0.013 | 1 | 1 |
| 12468-1107 |  | 0.0481 | Sy2 | D | < 0.004 | 0 | 1 |
| 12495-1308 |  | 0.0136 | Sy1 | D | < 0.004 | 0 | 1 |
| 12505-4121 | FAIRALL 315 | 0.0162 | Sy2 | D | < 0.004 | 0 | 1 |
| 12543-3006 |  | 0.0546 | Sy2 | D | 0.005 | 0 | 1 |
| 13044-2324 | NGC 4968 | 0.0093 | Sy2 | DS | 0.010 | 0 | 1 |
| 13059-2407 |  | 0.0141 | Sy2 | D | 0.171 | 0 | 1 |
| 13197-1627 | MCG -3-34- 63 | 0.0164 | Sy2 | DS | 0.026 | 0 | 1 |
| 13229-2934 | NGC 5135 | 0.0137 | Sy2 | S | < 0.005 | 1 | 1 |
| 13329-3402 | MCG -6-30- 15 | 0.0075 | Sy1 | DS | < 0.004 | 0 | 1 |
| 13357+0447* | NGC 5252 | 0.0231 | Sy2 | EN1 | < 0.006 | 1 | 1 |
| 13464-3003 | IC 4329A | 0.0144 | Sy1 | S | 0.009 | 0 | 1 |
| 13512-3731 | TOL 113 | 0.0520 | Sy1 | D | 0.009 | 0 | 1 |
| 14082+1347 | CGCG 74-129 | 0.0158 | Sy2 | D | < 0.004 | 0 | 1 |
| 14106-0258 | NGC 5506 | 0.0070 | Sy2 | DS | 0.048 | 1 | 0 |
| 14313+0540 | NGC 5674 | 0.0248 | Sy2 | E | < 0.006 | 1 | 0 |
| 14317-3237 |  | 0.0254 | Sy2 | D | < 0.004 | 0 | 1 |
| 14454-4343 | ESO 273-IG 4 | 0.0393 | Sy2 | N2 | < 0.003 | 0 | 1 |
| 14459-8248 |  | 0.1144 | Sy2 | D | 0.003 | 0 | 1 |
| 14557-2830 | ESO 448- G 10 | 0.0481 | Sy1 | D | < 0.004 | 0 | 1 |

TABLE 1-*Continued*

| IRAS name | alias | redshift | class | sample | PTI flux / Jy | no. of cuts | |
|---|---|---|---|---|---|---|---|
| (1) | (2) | (3) | (4) | (5) | (6) | (7) | (8) |
| 15015+1037 | MRK 841 | 0.0362 | Sy1 | DE | 0.055 | 0 | 1 |
| 15034+0353* | MRK 1392 | 0.0358 | Sy1 | N1 | < 0.006 | 0 | 1 |
| 15067+0913 | CGCG 077-021 | 0.0450 | Sy2 | D | 0.018 | 0 | 1 |
| 15091-2107 | | 0.0444 | Sy1 | D | 0.006 | 0 | 1 |
| 15184+0834 | | 0.0306 | Sy2 | D | < 0.005 | 0 | 1 |
| 15240+0046 | | 0.0508 | Sy2 | D | < 0.005 | 0 | 1 |
| 15288+0737 | NGC 5940 | 0.0339 | Sy1 | EN1 | < 0.005 | 1 | 0 |
| 15327+2340 | ARP 220 | 0.0182 | Sy2 | S | 0.017 | 0 | 1 |
| 15480-0344 | | 0.0303 | Sy2 | D | 0.024 | 0 | 1 |
| 15529+1920* | MRK 291 | 0.0352 | Sy1 | N1 | < 0.004 | 0 | 1 |
| 15599+0206 | | 0.1034 | Sy2 | D | 0.023 | 0 | 1 |
| 16062+1227 | MRK 871 | 0.0323 | Sy1 | DN1 | < 0.005 | 0 | 1 |
| 16277+2433 | MRK 883 | 0.0381 | Sy1 | N1 | < 0.005 | 0 | 2 |
| 18325-5926 | | 0.0192 | Sy2 | D | 0.023 | 0 | 1 |
| 18508-7815 | | 0.1610 | Sy1 | D | < 0.004 | 1 | 1 |
| 19169-5845 | ESO 141- G 55 | 0.0360 | Sy1 | D | < 0.003 | 0 | 1 |
| 19184-7404 | FAIRALL 513 | 0.0702 | Sy2 | N2 | < 0.004 | 0 | 1 |
| 19254-7245 | | 0.0615 | Sy2 | DN2 | 0.033 | 2 | 2 |
| 20162-5246 | FAIRALL 341 | 0.0163 | Sy2 | D | 0.004 | 0 | 1 |
| 20414-1054 | MRK 509 | 0.0340 | Sy1 | DS | < 0.004 | 0 | 3 |
| 20437-0259 | MRK 896 | 0.0262 | Sy1 | N1 | < 0.005 | 0 | 1 |
| 21052+0340 | MRK 897 | 0.0264 | Sy2 | N2 | 0.009 | 0 | 2 |
| 21299+0954 | II ZW 136 | 0.0621 | Sy1 | D | < 0.004 | 0 | 2 |
| 21538+0707 | MRK 516 | 0.0298 | Sy1 | N2 | < 0.003 | 0 | 2 |
| 21591-3206 | NGC 7172 | 0.0086 | Sy2 | S | 0.003 | 0 | 1 |
| 22017+0319 | | 0.066 | Sy2 | N2 | < 0.004 | 0 | 1 |
| 22045+0959 | NGC 7212 | 0.0260 | Sy2 | N2 | 0.030 | 0 | 1 |
| 22117-3903 | ESO 344- G 16 | 0.0394 | Sy1 | D | < 0.002 | 0 | 1 |
| 22340-1248 | MRK 915 far off | 0.0242 | Sy1 | D | 0.023 | 0 | 2 |
| 22377+0747 | UGC 12138 | 0.0246 | Sy1 | DEN1 | < 0.004 | 1 | 1 |
| 22570-2601 | | 0.0267 | Sy2 | N2 | < 0.003 | 0 | 2 |
| 22581-1311* | NGC 7450 | 0.0103 | Sy1 | N1 | < 0.006 | 1 | 0 |
| 23016+2221 | MRK 315 | 0.0385 | Sy1 | DN1 | < 0.006 | 0 | 1 |
| 23027-0004 | UGC 12348 | 0.0253 | Sy2 | D | < 0.002 | 0 | 2 |
| 23069-4341 | NGC 7496 | 0.0500 | Sy2 | N2 | 0.007 | 0 | 1 |
| 23156-4238 | NGC 7582 | 0.0053 | Sy2 | S | < 0.005 | 1 | 0 |
| 23161-4230 | NGC 7590 | 0.0050 | Sy2 | S | < 0.003 | 0 | 1 |
| 23163-0001 | MRK 530 | 0.0290 | Sy1 | EN1 | 0.011 | 1 | 2 |
| 23265+0315* | NGC 7682 | 0.0170 | Sy2 | E | 0.040 | 1 | 1 |
| 23534+0714* | MRK 541 | 0.0391 | Sy1 | N1 | < 0.004 | 0 | 1 |
| 23598+0304* | MRK 543 | 0.0255 | Sy1 | N1 | < 0.005 | 0 | 1 |

NOTES.-Col 1: IRAS, or IRAS-like, name. '*' indicates an IRAS non-detection; col 2: catalog names; col 3: redshifts from the literature; col 4: spectral classification from the literature; col 5: samples in which each galaxy appears. D - de Grijp et al. 1987, E - Edelson 1987, N1 - Norris et al. 1990 optically selected sample, N2 - Norris et al. 1990 FIR-selected sample, S - Spinoglio & Malkan 1989; col 6: 2295-MHz PTI flux density in Jy. Upper limits are quoted at five times the rms noise in the fringe-frequency spectrum; col 7: the number of observations made at 1665 MHz. These measurements were converted to 2295 MHz assuming $s \sim n^{-0.7}$; col 8: the number of observations made at 2295 MHz.

TABLE 2
OBSERVING DATES AND TELESCOPE CONFIGURATIONS

| Date | Freq 1 MHz | Freq 2 MHz | Pol 1 | Pol 2 | $T_{sys}$ / Jy | | Integration sec | clock | |
|---|---|---|---|---|---|---|---|---|---|
| | | | (IEEE convention) | | Tid | PKS | | Tid | Pks |
| 10 - 14 Dec 87 | 1662.0 | 1669.0 | LHC | LHC | 45 | 100 | 512 | H-maser | Rb |
| 02 - 03 Aug 88 | 2286.5 | 2293.5 | RHC | RHC | 45 | 100 | 512 | H-maser | Rb |
| 21 - 24 Nov 88 | 2290.0 | 2290.0 | LHC | RHC | 15 | 100 | 512 | H-maser | Rb |
| 26 - 27 Apr 90 | 2286.5 | 2293.5 | RHC | RHC | 30 | 90 | 1024 | H-maser | Rb |
| 01 Jul 91 | 2286.5 | 2293.5 | RHC | RHC | 30 | 90 | 1024 | H-maser | Rb |
| 26 - 27 Jul 91 | 2286.5 | 2293.5 | RHC | RHC | 30 | 90 | 1024 | H-maser | Rb |
| 24 Sep 91 | 2286.5 | 2293.5 | RHC | RHC | 30 | 90 | 1024 | H-maser | H-maser |
| 26 Oct 91 | 2286.5 | 2293.5 | RHC | RHC | 30 | 90 | 1024 | H-maser | H-maser |

REFERENCES

Allen, D. A., Norris, R. P., Meadows, V. S., & Roche, P. F. 1991, MNRAS, 248, 528

Antonucci, R. R. J, & Miller, J. S. 1985, ApJ, 297, 621

Baars, J. W. M., Genzel, R., Pauliny-Toth, I. I. K., & Witzel, A. 1977, A&A, 61, 99

Barthel, P. D., 1989, ApJ, 336, 606

Bothun, G. D., & Rogers, C. 1992, AJ, 103, 1484

Bregman, J. N. 1990, A&AR, 2, 125

Burstein, D., & Lebofsky, M. J. 1986, ApJ, 301, 683

Cohen, M. H., et al. 1975, ApJ, 201, 249

Dahari, O., & De Robertis, M. M. 1988, ApJS, 67, 249

de Grijp, M. H. K., Miley, G. K., & Lub, J. 1987, A&AS, 70, 95

Edelson, R. A. 1987, ApJ, 313, 651

Evans, I. N., Ford, H. C., Kinney, A. L., Antonucci, R. R. J., Armus, L., and Caganoff, S. 1991, ApJ, 369, L27

Feigelson, E. D., & Nelson, P. I. 1985, ApJ, 293, 192

Fitt, A. J., Howarth, N. A., Alexander, P., & Lasenby, A. N. 1992, MNRAS, 255, 146

Glantz, S. A. 1992, *Primer of Biostatistics,* 3rd ed., (New York: McGraw-Hill)

Haniff, C. A., Wilson, A. S., & Ward, M. J. 1988, ApJ, 334, 104

Hoel, P. G., 1966, *Elementary Statistics,* 2nd ed., (New York: Wiley), 166-167

*IRAS Point Source Catalog,* 1985, Joint IRAS Science Working Group (Washington, DC: US Government Printing Office)

Khachikyan, É. Ya., & Weedman, D. W. 1971, Astrophys., 7, 231

Krolik, J. H., & Begelman, M. C. 1986, ApJ, 308, L55

Krolik, J. H., & Begelman, M. C. 1988, ApJ, 329, 702

Krolik, J. H., & Lepp, S. 1989, ApJ, 347, 179

Kukula, M. J., Ghosh, T., Pedlar, A., Schilizzi, R. T., Miley, G. K., de Bruyn, A. G., & Saikia, D. J. 1993, MNRAS, 264, 893

La Valley, M. P., Isobe, T., & Feigelson, E. D. 1992, BAAS, 24, 839

Lawrence, A. 1987, PASP, 99, 309

Miller, J. S., & Goodrich, R. W. 1990, ApJ, 355, 456

Netzer, H., & Laor, A. 1993, ApJ, 404, L51

Norris, R. P., Allen, D. A., Sramek, R. A., Kesteven, M. J., & Troup, E. R. 1990, ApJ, 359, 291

Norris, R. P., Kesteven, M. J., Wellington, K. J., & Batty, M. J. 1988, ApJS, 67, 85

Norris, R. P., Roy, A. L., Allen, D. A., Kesteven, M. J., Troup, E. R., & Reynolds, J. E. 1992, in *Relationships Between Active Galactic Nuclei and Starburst Galaxies*, ed. Alexei V. Filippenko, ASP Conference Series, Vol. 31, 71

Osterbrock, D. E. 1987, in *Spectroscopy of Astrophysical Plasmas*, ed. A. Dalgarno, & D. Layzer, (Cambridge University Press), 59

Osterbrock, D. E., & Koski, A. T. 1976, MNRAS, 176, 61p

Pedlar, A., Dyson, J. E., & Unger, S. W. 1985, MNRAS, 214, 463

Penston, M. V., & Pérez, E. 1984, MNRAS, 211, 33p

Pounds, K. A., Nandra, K., Stewart, G. C., George, I. M., & Fabian, A. C. 1990, Nature, 344, 132

Rao, A. R., Singh, K. P., & Vahia, M. N. 1992, MNRAS, 255, 197

Reynolds, J. E. 1992, private communication

Roy, A. L., Norris, R. P., Kesteven, M. J., Reynolds, J. E., & Troup, E. R. 1994, in preparation

Soifer, B. T., Houck, J. R., & Neugebauer, G. 1987, ARA&A, 25, 187

Spinoglio, L., & Malkan, M. A. 1989, ApJ, 342, 83

Terlevich, R., & Melnick, J. 1985, MNRAS, 213, 841

Turner, T. J., et al. 1993, ApJ, 407, 556

Ulvestad, J. S., & Wilson, A. S. 1989, ApJ, 343, 659

Ulvestad, J. S., Wilson, A. S., & Sramek, R. A. 1981, ApJ, 247, 419

Unger, S. W., Pedlar, A., Booler, R. V., & Harrison, B. A. 1986, MNRAS, 219, 387

Veilleux, S. 1991, ApJ, 369, 331

Véron-Cetty, M. -P., & Véron, P. 1991, *A Catalogue of Quasars and Active Nuclei* 5th ed., ESO Scientific Report No. 10

Whittle, M., Haniff, C. A., Ward, M. J., Meurs, E. J. A., Pedlar, A., Unger, S. W., Axon, D. J., & Harrison, B. A. 1986, MNRAS, 222, 189

Whittle, M., Pedlar, A., Meurs, E. J. A., Unger, S. W., Axon, D. J., & Ward, M. J. 1988, ApJ, 326, 125

Wilson, A. S., & Willis, A. G. 1980, ApJ, 240, 429

Wilson, A. S. 1993, private communication.

Zheng, W. 1992, ApJ, 385, 127

FIGURE CAPTIONS

FIG. 1.-Redshift distributions of Sy1s and Sy2s compared, (a) for the FIR-selected Seyferts, and (b) for the combined optical+mid-IR-selected sample. For each Seyfert type, two curves are shown which represent the fraction of the sample that has a redshift greater than a given value. The width of the region shows the 95% confidence interval for the redshift distribution (the curves are the cumulative form of the 95% confidence intervals of the Kaplan-Meier estimator). There is no significant difference between the redshift distributions of the two Seyfert types in the FIR-selected sample, but there is a clear difference in the optical+mid-IR-selected sample.

FIG. 2.-[OIII] luminosity distribution of de Grijp et al. FIR-selected Seyfert 1s and Seyfert 2s compared. The 95% confidence intervals of the Kaplan-Meier estimator, in cumulative form, are shown for the two Seyfert types separately, as in Figure 1. The distributions are similar, but there may be a tendency for Sy2s to be lower luminosity.

FIG. 3.-PTI detection rates for the FIR-selected Seyfert sample and for the combined optical and mid-IR-selected Seyfert sample, separated into Seyfert types 1 and 2. The ratios indicate the number detected / total in sample. The detection rates are different for the two Seyfert types.

FIG. 4.-PTI detection rates for the de Grijp et al. FIR-selected Seyfert sample, and for the Norris et al. FIR-selected sample, separated into Seyfert types 1 and 2. The ratios indicate the number detected / total in the sample. The detection rates are different for the two Seyfert types.

FIG. 5.-Schematic diagram illustrating the two proposed mechanisms which explain why we see radio cores in Sy2s but not in Sy1s. The NLR clouds are optically thick at 2.3 GHz, and so in Sy1s the core cannot be seen and only radio emission from NLR clouds on one side of the nucleus is visible. However, in Sy2s the radio emission from both regions is visible.

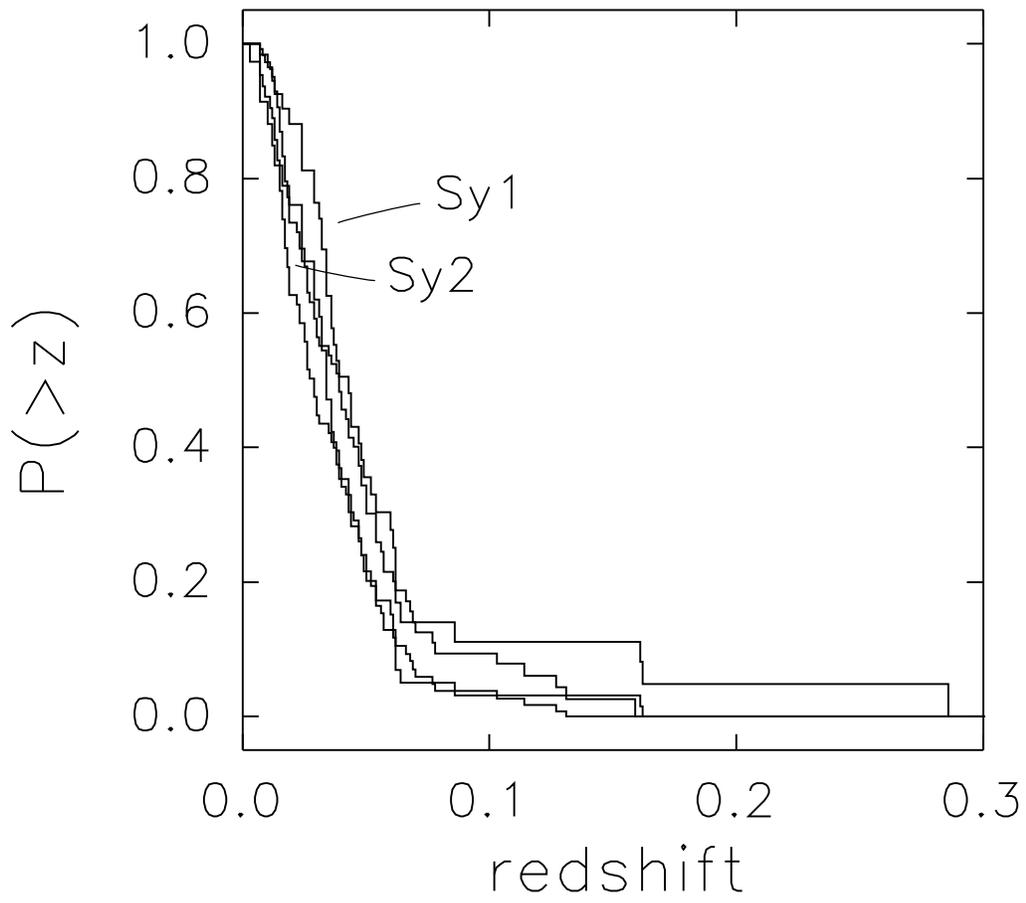

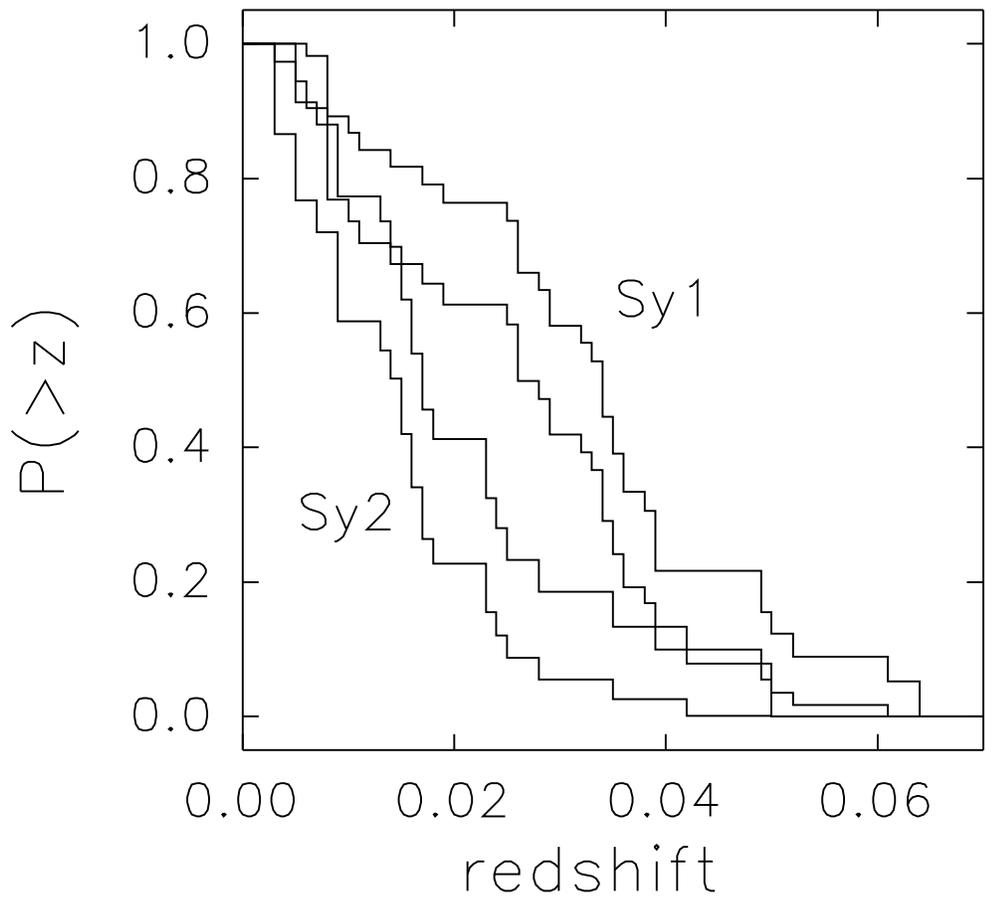

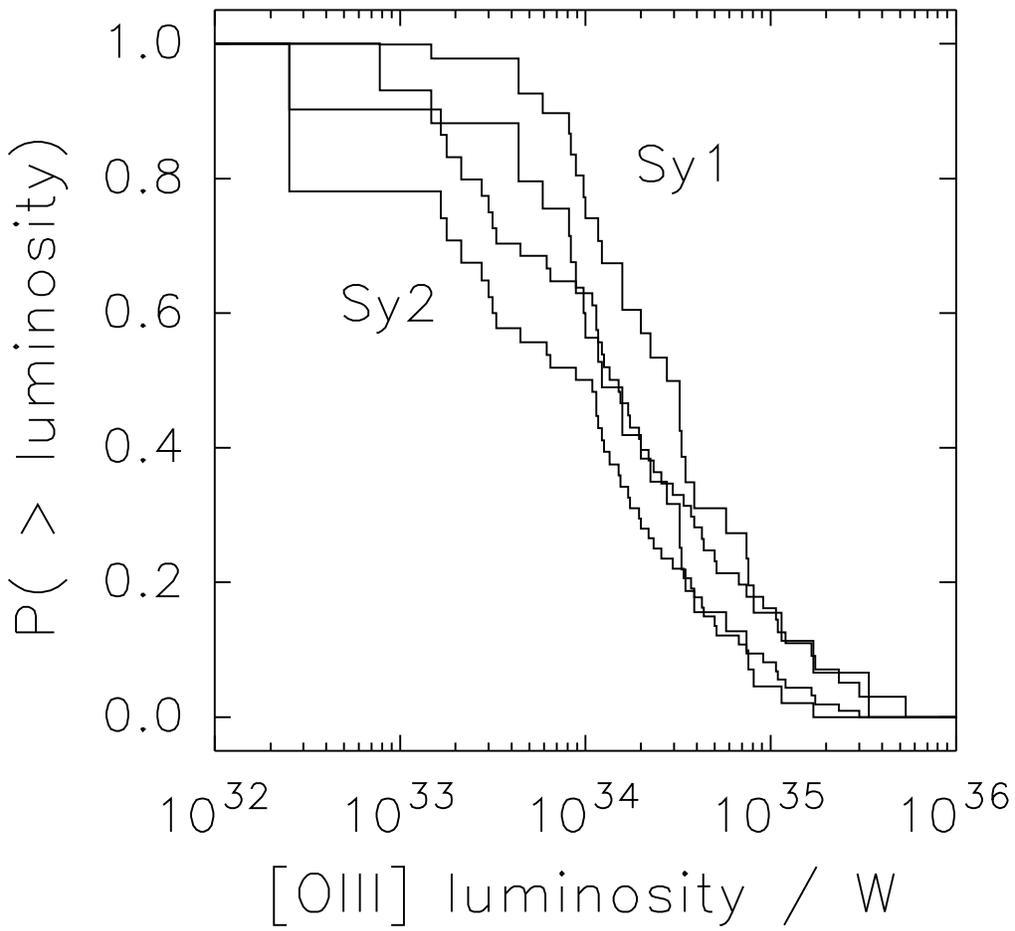

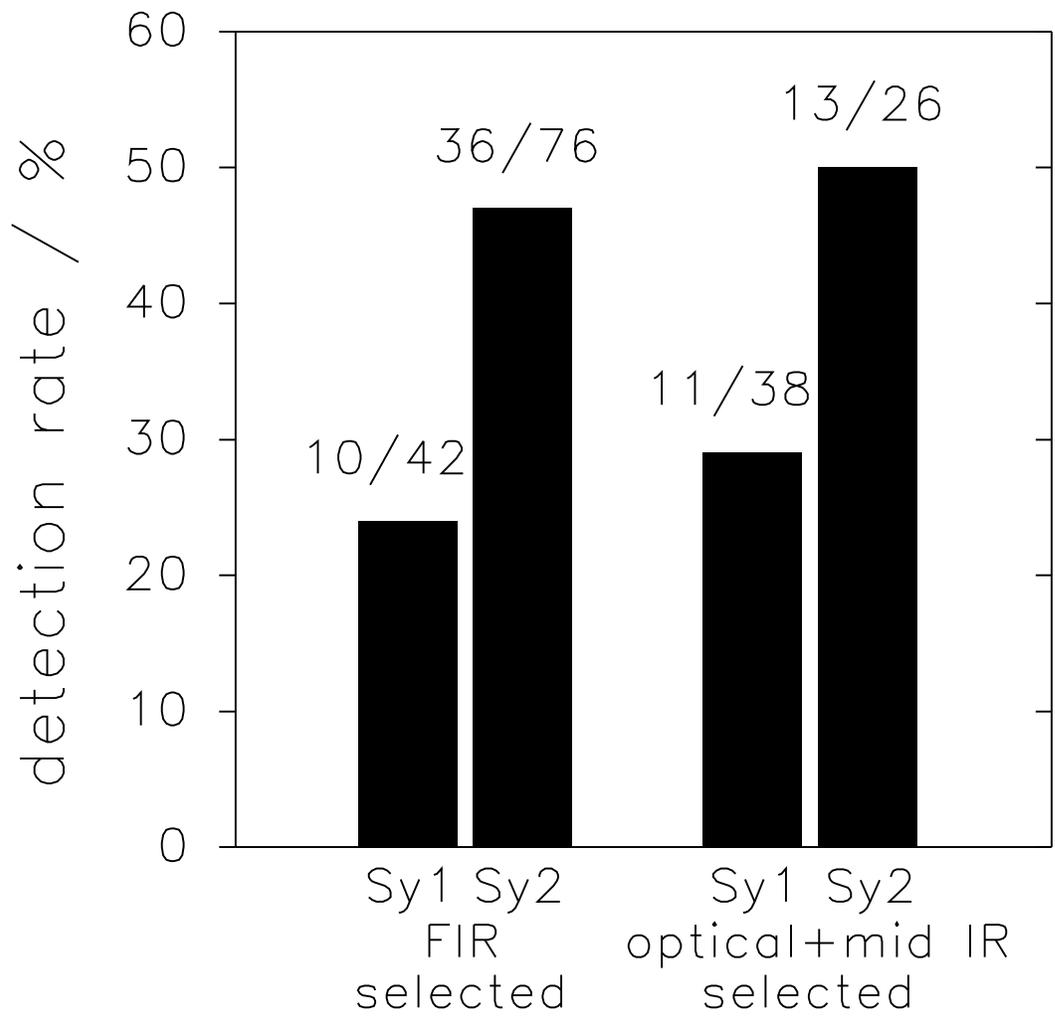

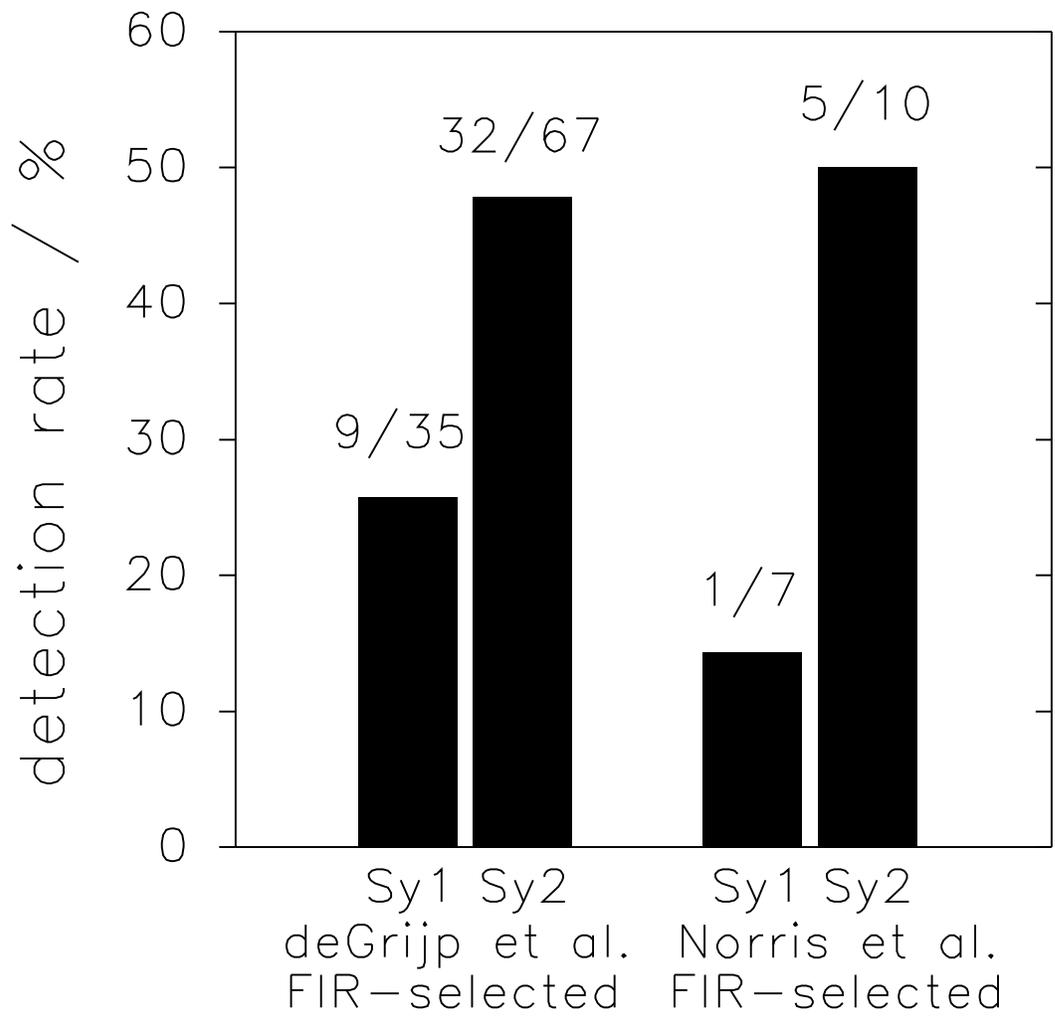

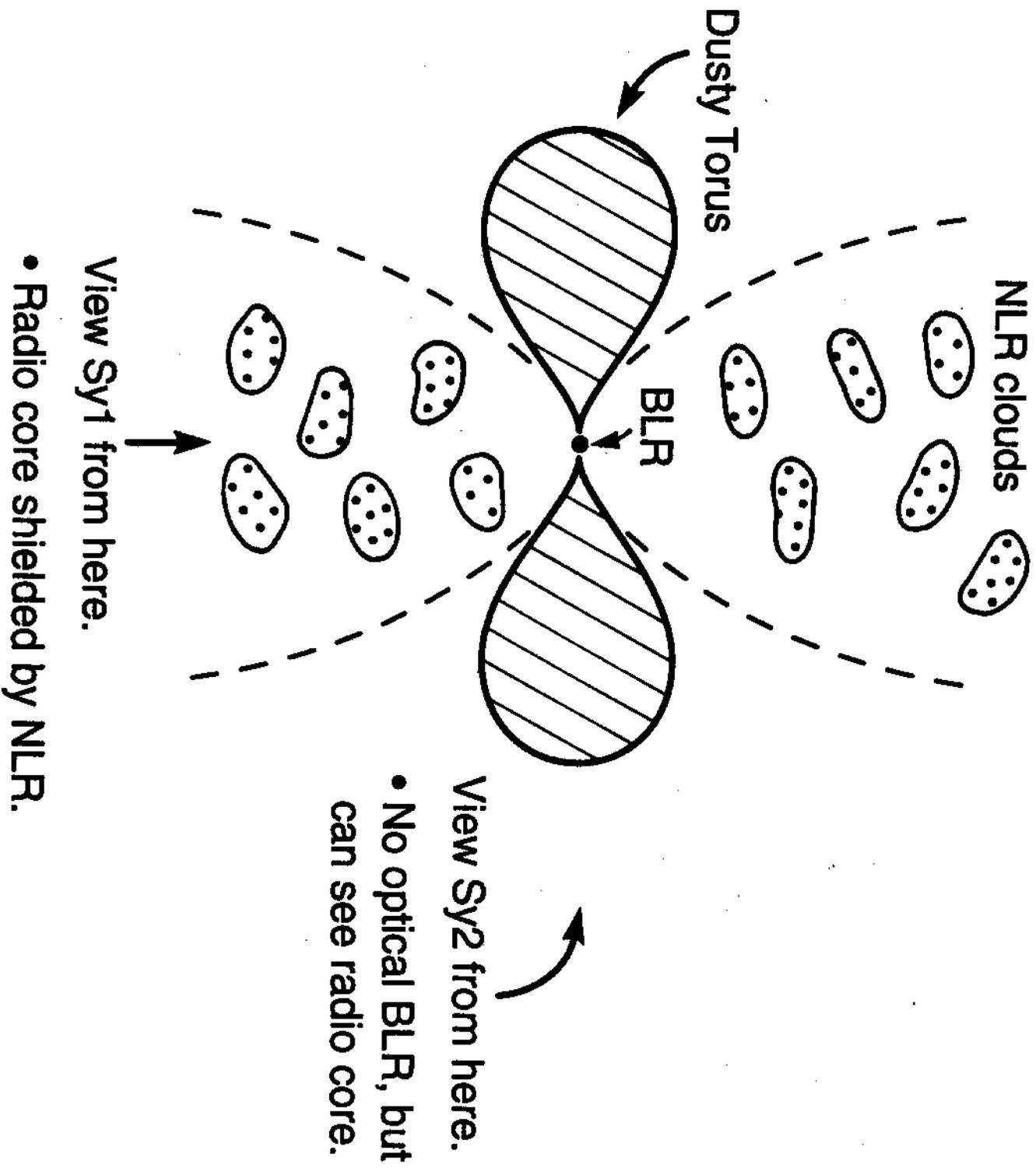

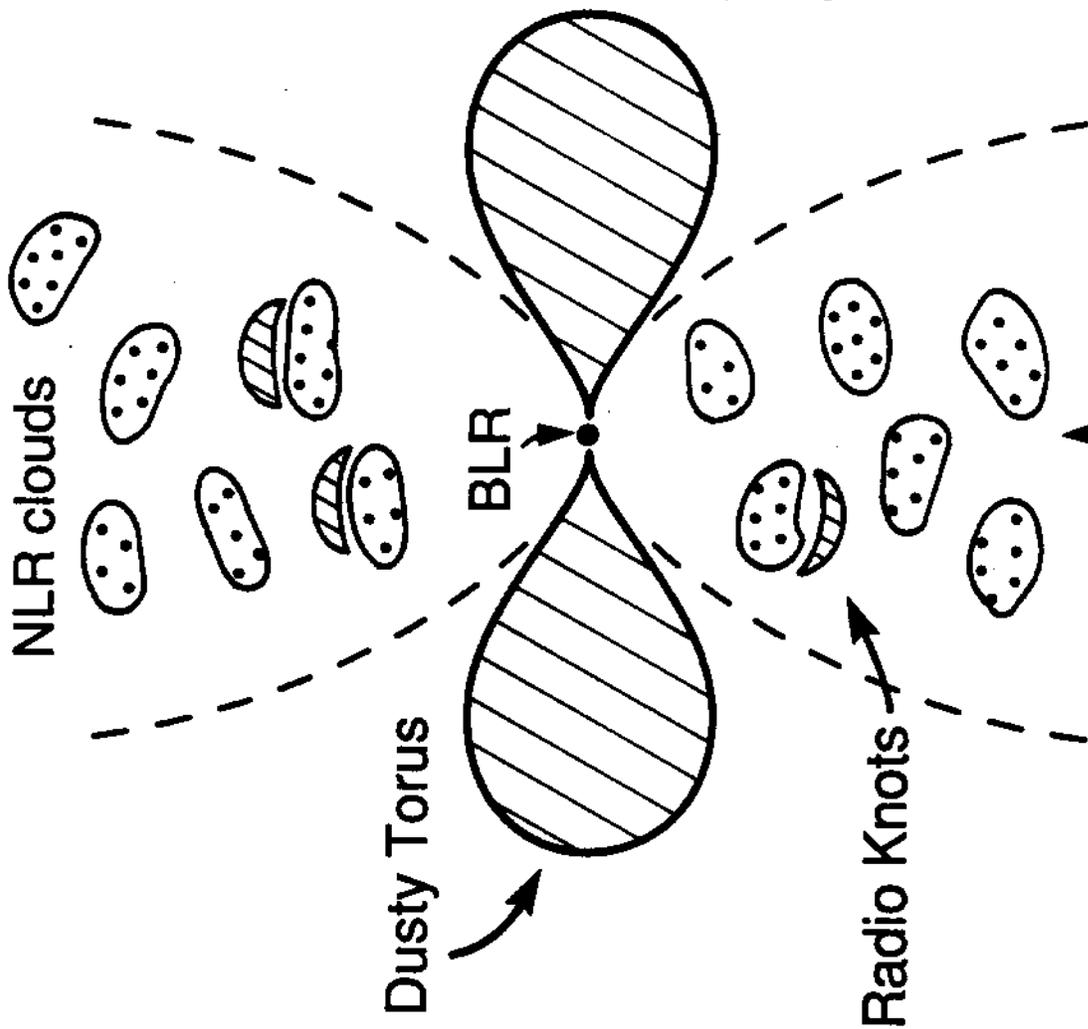